# Superradiance of a subwavelength array of independent classical nonlinear emitters

N. E. Nefedkin,[1,2,3] E. S. Andrianov,[1,2] A. A. Zyablovsky,[1,2] A. A. Pukhov,[1,2,3] A. P. Vinogradov,[1,2,3] and A. A. Lisyansky[4,5]

[1]*Moscow Institute of Physics and Technology, 9 Institutskiy per., Dolgoprudny 141700, Russia*
[2]*All-Russia Research Institute of Automatics, 22 Sushchevskaya, Moscow 127055, Russia*
[3]*Institute for Theoretical and Applied Electromagnetics, 13 Izhorskaya, Moscow 125412, Russia*
[4]*Department of Physics, Queens College of the City University of New York, Queens, NY 11367, USA*
[5]*The Graduate Center of the City University of New York, New York, New York 10016, USA*

We suggest a mechanism for the emergence of a superradiance burst in a subwavelength array of nonlinear classical emitters. We assume that the emitters interact via their common field of radiative response and that they may have an arbitrary distribution of initially phases. We show that only if this distribution is not uniform, a non-zero field of radiative response arises leading to a superradiance burst. Although this field cannot synchronize the emitters, it forces fast oscillations of a classical nonlinear emitter to have long-period envelopes. Constructive interference in the envelopes creates a large dipole moment of the array which results in a superradiance pulse. The intensity of the superradiance is proportional to the squared number of the emitters, which envelopes participate in the fluctuation.

# I. INTRODUCTION

Superradiance (SR) is an enhancement of collective spontaneous radiation of an array of emitters interacting with a common light field. For a subwavelength array of quantum emitters this phenomenon was predicted by Dicke.[1] In the Dicke model, once $N$ emitters are excited, they radiate a pulse after a time $t_0 \sim \log N / N$. The duration of the SP pulse is smaller than the radiation time of a single emitter by the factor of 1/$N$. The intensity of the radiation is proportional to $N^2$. The Dicke model assumes that emitters interact with each other via the common field of their radiative response. The Dicke approach only considers the dynamics of the population inversion of quantum emitters without discussing their phases. The intensity of the SR radiation is defined as the time derivative of the total inversion. This is correct only if all emitters have the same phase. Unfortunately, the formal Dicke's approach does not allow one to elucidate the physical mechanism of phasing.

Phenomena similar to SR have been observed experimentally only in extended systems with size greatly exceed the optical wavelength, at least in one dimension.[2-4] The results of these experiments are not unambiguous because in extended systems a significant contribution to radiation may arise due to waves originated at one end and then radiated at the other.[4,5] Thus, it is hard to separate contributions due to SR and the stimulated emission.[2-4]

At the same time, the theoretical considerations have focused on increasing of the accuracy of the description of a quantum system.[4,6,7] E.g., an approach based on the density matrix formalism [4,6,8] makes the obtained result more rigorous but does not expose its physics. In any case, it has commonly been recognized that for SR to occur identical quantum emitters have to be in the special "Dicke state."[4]

SR has generally been considered to be a purely quantum mechanical phenomenon.[5,9] However, Vainshtein and Kleev[10] showed that similar to quantum emitters, an ensemble of classical nonlinear emitters interacting via their common radiation field may radiate a SR pulse.[10,11] The Vainshtein-Kleev model assumes that initially excited classical emitters, which are not pumped further, have random distribution of phases. Computer simulation[10] showed that classical linear emitters do not become superradiant while in case of cubic nonlinearity depending on the initial phase distribution realization either one pulse or a sequence of pulses may arise. This is similar to quantum oscillators, which response to a self-consistent field is always nonlinear [12-14].

Despite extensive studies, the physical mechanism for SR of classical emitters is still unclear.[15,16] It is well known [17,18] that a system of $N$ classical emitters oscillating in phase loses energy $N$ times faster than a single oscillator. This can be interpreted as SR. In this approach, the main theoretical question is how excited emitters initially having different phases evolve into in-phase oscillations. Such a mechanism exists for nonlinear auto-oscillating systems with continuous excitation. This is well-known synchronization by a driving force. In this case, all auto-oscillations have the phase and frequency of the external force. In Refs. 19-21, the field of the response radiation of a collection of auto-oscillating systems (spasers) was considered as a synchronizing factor. The radiation intensity was predicted to be proportional to the square of the number of emitters. However, since the synchronized auto-oscillating process is stationary, it cannot give rise to a SR pulse.

In this paper, we suggest a mechanism for the emergence of a SR pulse in an ensemble of classical nonlinear dipole emitters. Following Dicke[1] and Vainshtein-Kleev,[10] we assume that each dipole is in the total field of radiative response of the whole system. This field is produced by all dipoles and depends on their phase distribution. We show that this field may arise only due to a fluctuation in the dipole phase distribution which is initially uniform. This field causes a modulation of the fast oscillations of dipoles with a periodic envelope. The frequency of the envelope is determined by the initial phase of the dipole oscillation. This frequency is much smaller than the frequency of the dipole oscillation. SR arises due to constructive interference in long-period envelopes of fast oscillations which causes an increase of the amplitude of the oscillation of the total dipole moment of the system. The lifetime of the large amplitude oscillations is of the order of the envelope period which is much greater than the period of fast oscillations. During this lifetime, the dipoles superradiate.

## II. DYNAMICS OF INTERACTING CLASSICAL NONLINEAR DIPOLES

Following Ref. 10, we consider a system of oscillating dipoles placed in a subwavelegth volume $V \ll \lambda^3$. We assume that the energy of dipoles oscillating with the frequency $\omega$ is much greater than $\hbar\omega$, so that the classical theory is applicable. We use the model suggested in Refs. 10, 11 to describe the dynamics of these oscillators.

The field of the each oscillator can be expressed via the Hertz vector:[22, 23]

$$\mathbf{\Pi} = -\frac{1}{r}\mathbf{d}\left(t - \frac{r}{c}\right), \tag{1}$$

where $\mathbf{d}$ is the dipole moment of an oscillator and $\mathbf{r}$ is the distance from the oscillator to the observation point. The Fourier component the vector $\mathbf{\Pi}$ has the form

$$\mathbf{\Pi}_\omega = -\frac{\mathbf{d}_\omega \exp(ik_0 r)}{r}, \tag{2}$$

where $k_0 = \omega/c$. Since we are interested in the field at small distances, $k_0 r \ll 1$, we expand $\mathbf{\Pi}_\omega$ into a series

$$\mathbf{\Pi}_\omega = -\mathbf{d}_\omega \left(1 + ik_0 r - k_0^2 r^2/2 - ik_0^3 r^3/6 + ...\right)/r. \tag{3}$$

Assuming that all dipoles are directed along the z-axis we find the z-component of the electric field:

$$E_\omega^z = -k^2 \Pi_\omega^z - \Delta \Pi_\omega^z = d_\omega \left( \frac{3z^2 - r^2}{r^5} + \frac{k_0^2(r^2 + z^2)}{2r^3} + \frac{2ik_0^3}{3} + ... \right). \quad (4)$$

Since $d = ez$ is the dipole moment of a charge oscillating along the z-axis and $i^n \partial^n z / c^n \partial t^n$ can substitute for $zk_0^n$, we finally obtain

$$E_z = e \left[ \frac{1 - 3\cos^2 \alpha}{cr^2} - \frac{1 + \cos^2 \alpha}{2c^2 r} \ddot{z}(t) + \frac{2}{3c^3} \dddot{z}(t) + ... \right], \quad (5)$$

where $z(t)$ is an instantaneous position of the oscillator (dipole) and $\alpha$ is the angle between **r** and the z-axis.[11, 22, 23] Since the oscillators are confined in a subwavelength volume, the retardation effects can be neglected, we omit, therefore, the terms with derivatives higher than $\dddot{z}$ in Eq. (5). In this equation, the first term corresponds to a quasistatic Coulomb field and the second term, proportional to $1/r$, describes the induction field. These fields suppress SR.[22] Below, we neglect this effect assuming that emitters are positioned either along a circle[4] or form an ideal cubic lattice. In both cases these fields turn to zero due to the symmetry of the problem. The third term,

$$E_z = (2e/3c^3) \dddot{z}(t), \quad (6)$$

expresses the field of the radiative response. Due to its active nature, it is in an anti-phase with the dipole current. The equation of motion of the oscillator due to this field has the form

$$\ddot{z} + \omega^2 z = t_e \dddot{z}(t), \quad (7)$$

where $t_e = (2e^2)/(3mc^3) = (2r_0)/3c = 6.27 \cdot 10^{-24} c$, and $r_0 = e^2/mc^2 = 2.82 \cdot 10^{-13}$ cm is the classical radius of an electron. Using the smallness of $t_e$ to estimate $\dddot{z}$ we can omit the right-hand side in Eq. (7): $\ddot{z} + \omega^2 z = 0$. Then, in the first order in $t_e$ we obtain

$$\ddot{z} + \omega^2 z = -t_e \omega^2 \dot{z}(t) \quad (8)$$

For one dipole, the field of the radiative response is defined by Eq. (6). For $\omega r/c \ll 1$, field (6) does not depend on a distance and for N dipoles we have

$$E_z = (2e/3c^3) N \langle \dddot{z} \rangle, \quad z = N^{-1} \sum_{n=1}^{N} z_n. \quad (9)$$

Then for an ensemble of N nonlinear dipoles we obtain a system of equations

$$\ddot{z}_k + \omega^2 z_k + \mu z_k^3 = -N\nu \langle \dot{z} \rangle, \tag{10}$$

where $\nu = 2e^2\omega^2/3mc^2$.

Now, instead of real quantities $z_k$ we introduce complex dimensionless variables, envelopes $c_k$:

$$z_k(t) = a\left(c_k(t)e^{-i\omega t} + c_k^*(t)e^{i\omega t}\right)/2 = a\mathrm{Re}\left(c_k(t)e^{-i\omega t}\right), \tag{11}$$

where $a$ is an initial amplitude of oscillations and $c_k(0) \sim e^{i\varphi_k}$. Note that all phase detunings are included into the complex amplitude. Fast oscillations filling the envelope are in phase. Instead of one real variable we introduced two variables, $\mathrm{Re}\,c_k$ and $\mathrm{Im}\,c_k$. To remove an extra variable we can impose an additional condition:[11, 19]

$$\dot{c}_k e^{-i\omega t} + \dot{c}_k^* e^{i\omega t} = 0. \tag{12}$$

By averaging Eq. (10) and taking into account smallnesses of the attenuation and nonlinearity we obtain

$$\dot{c}_k + i\chi\omega\left(|c_k|^2 - 1\right)c_k = -\frac{1}{2}N\nu\langle c \rangle, \quad \langle c \rangle = \frac{1}{N}\sum_{k=1}^{N} c_k, \tag{13}$$

where $\chi = 3\mu a^2/8\omega^2$ is the coefficient of non-isochronism.[11] By using dimensionless parameters,

$$\tau = t/\tau_N = N\nu t/2, \quad \tau_N = 2/(N\nu), \quad \theta = 2\chi\omega/(N\nu), \tag{14}$$

equations (13) for slowly varying amplitudes can be simplified

$$\frac{dc_k}{d\tau} + i\theta\left(|c_k|^2 - 1\right)c_k = -\langle c \rangle, \tag{15}$$

where the parameter $\theta$ may be either positive or negative. Equation (15) with $\theta = 0$ describes linear oscillators.

### III. DYNAMICS OF SLOWLY VARYING AMPLITUDES

Let us consider the case of randomly distributed amplitudes.[10] We illustrate it for the values of $\theta = 10$ and the number of dipoles $N = 250$. At the moment of time $\tau = 0$, $|c_k(0)| = 1$ and phases of oscillations of all particles are random. The result of numerical calculations of the average intensity, $\left|\langle c_k \rangle\right|^2$, and the average energy, $\langle |c_k|^2 \rangle$, on time are shown in Fig. 1.

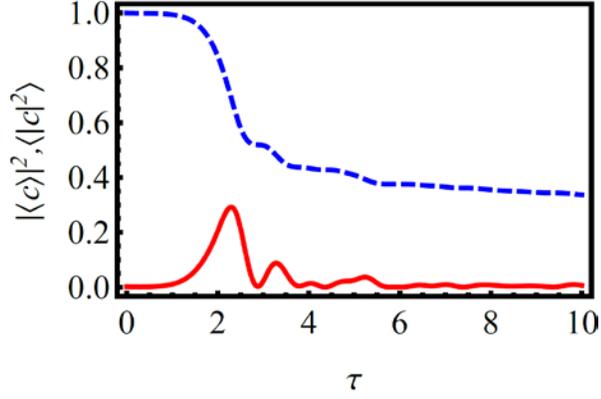

FIG 1. The results of numerical calculations of the average intensity, $\left|\langle c_k \rangle\right|^2$, (the dashed line) and the average energy, $\langle |c_k|^2 \rangle$, (the solid line) on time $\tau$.

The dependencies shown in Fig. 1 are in agreement with those in Ref. 10 and qualitatively correspond to the usual picture of Dicke SR in which the first large peak of the intensity has the duration of $\tau_s \tau_N \sim 1/N$, and its delay time is $\tau_0 \sim \log N$ (see Fig. 2). Note that the dimensional delay time has the same dependency on the number of emitters as in the Dicke model, $t_0 = \tau_N \tau_0 \sim \log N / N$.

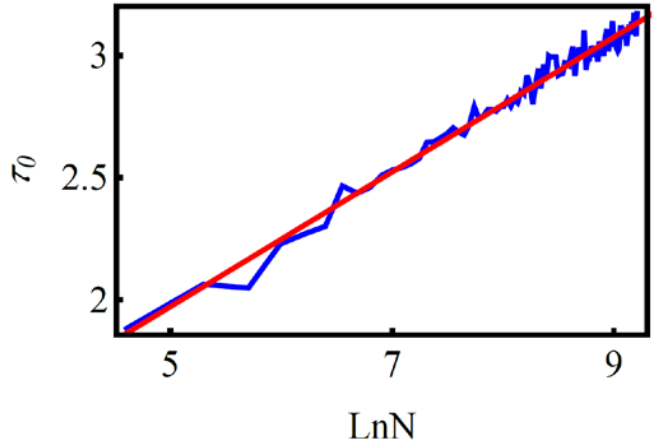

FIG. 2. The dependence of the delay time $\tau_0 = t_0 / \tau_N$ on the number of dipoles for the random initial phase distribution in the interval $[-\pi, \pi]$. The approximation line $\alpha_{\tau_0} \log N$ with $\alpha_{\tau_0} \approx 0.5$ is shown in red.

The results of our numerical experiment also show a quantitative difference from the dependencies predicted by Dicke. In the Dicke model, the intensity of the SR peak is proportional to $N^2$, while our numerical calculations do not show a power dependence of the intensity on the number of emitters (see Fig. 3).

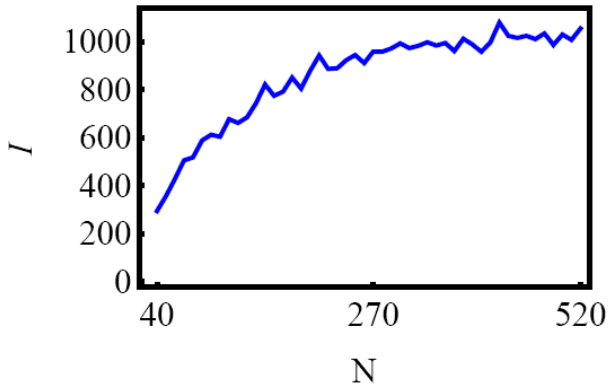

FIG. 3. The dependence of the peak intensity of radiation on the number of dipoles for the random distribution of initial phases in the interval $[-\pi, \pi]$.

In order to clarify the cause of the discrepancy between the quantum Dicke model and the classical model of Ref. [10] we simplify the latter model by using the mean field (MF) approximation. This allows us to reveal the mechanism of claster syncronization.

## IV. MEAN FIELD MODEL OF SUPERRADIATIVE DIPOLES

The average value of the dipole $\langle c \rangle = N^{-1} \sum_{k=1}^{N} c_k$, in the right hand side of Eq. (15), determines the MF acting on all dipoles. For the random distribution of dipole initial phases, one can expect that $\langle c \rangle = 0$. However, since we deal with a finite number of dipoles, $N$, then $\langle c^2 \rangle \sim N$. In each realization of initial phases, the average dipole moment $\langle c \rangle \sim \sqrt{N}$ is never zero and dipoles are in a nonzero MF. If we choose a uniform distribution of initial phases, then $\langle c \rangle = 0$ and a SR pulse does not arise. Therefore, a fluctuation of the initial phase distribution which creates $\langle c \rangle \neq 0$ is necessary for SR.

For the sake of simplicity, we presume that the MF remains constant until the system superradiate. After that, dipoles lose their energy and the field of each of them turns to zero.

After $\langle c \rangle$ is replaced by a constant, the emitters become independent and the dynamics of each of them is described by the equation

$$\frac{dc}{d\tau} + i\theta\left(|c|^2 - 1\right)c = -E, \qquad (16)$$

where $E$ is a complex-valued constant corresponding to the value of the MF.

The dynamics of each dipole varies due the differences in dipole initial conditions. This difference is not only in a simple phase shift. Nonlinearity of the emitters results in more complicated behavior. Indeed, Eq. (10) is the Duffing equation whose solution may be a high frequency oscillations modulated by a low frequency envelope.[24] In this case, Eq. (16) describing the envelope evolution has a periodic solution with a period determined by initial conditions. This is illustrated by Fig. 4 which shows that periods of oscillations of dipoles depend on initial phases.

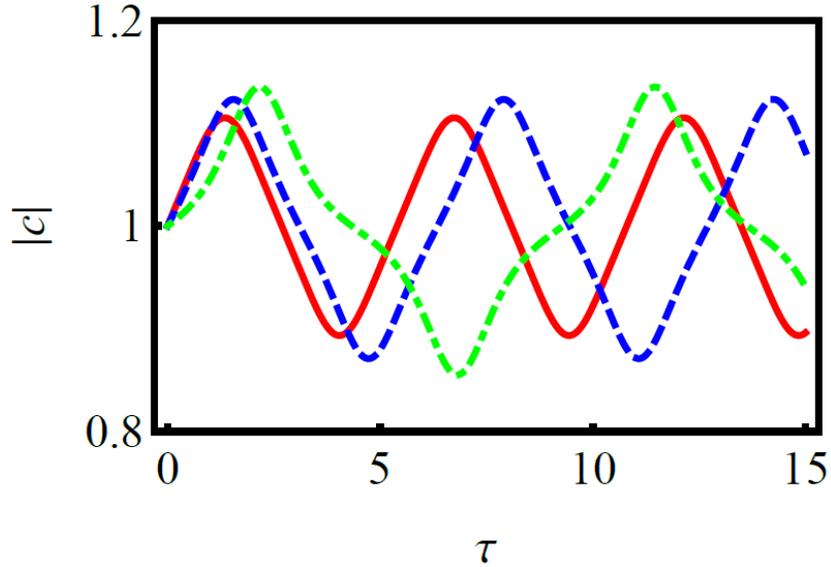

FIG. 4. Dependencies of periods of dipole oscillations on time. All dipoles have the same initial amplitude but different phases: Arg($c$) = 0 (the red line), Arg($c$) = $\pi$/10 (the blue line), and Arg($c$) = $\pi$/5 (the green line). The dipoles are at the same MF, $\tau = t/\tau_N$.

If the initial distribution of phases is uniform, the classical model[10] predicts $E = 0$ and no SR. In the MF model, $E$ is a parameter. If we put $E = 0$, we also obtain that there is no SR. On the other hand, for $E \neq 0$ even for the uniform initial distribution of phases, SR arises (see Fig. 5). Note that in this approximation, all the dipoles do not interact with each other but the sum of the solutions of Eq. (16) brings about the radiation intensity characteristic for SR, as shown in Fig. 5.

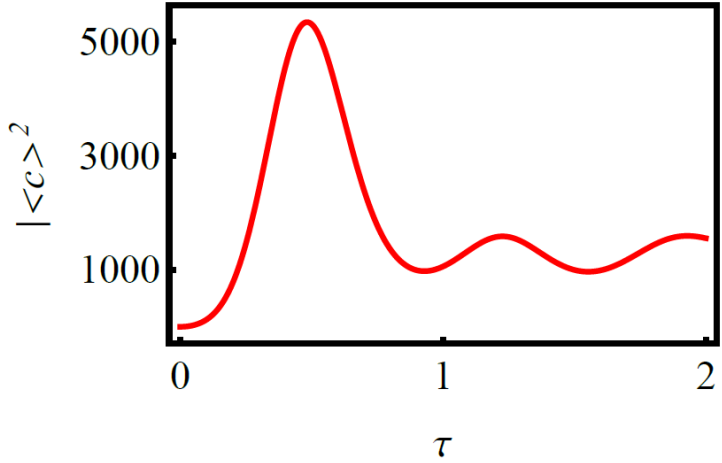

FIG. 5. The intensity of radiation of the system of non-interacting dipoles with $E \neq 0$.

Let us represent amplitudes of envelopes and the field as $c_i = A_i \exp(i\varphi_i)$ and $E = E_0 \exp(i\psi)$, respectively. At the initial moment, we neglect the amplitude change compared to the phase change. Then $A_i$ can be considered as constants. Now, Eq. (16) can be reduced to the equation for the phase dynamics:

$$\frac{d\varphi_i}{d\tau} + \theta\left(A_i^2 - 1\right) = E_0 \sin(\varphi_i - \psi) / A_i. \quad (17)$$

At some points, the phase derivative in Eq. (17) turns to zero. These points are determined by the equation $\theta\left(A_i^2 - 1\right) = E_0 \sin(\varphi_i - \psi) / A_i$. Fig. 4 shows that the amplitudes $A_i$ are close to unity. Therefore, $\theta\left(A_i^2 - 1\right) \approx 0$ and Eq. (17) may be simplified as $\dot{\phi}_i = E_0 \sin\phi_i$ with $\phi_i = \varphi_i - \psi$. This equation has two fixed points in which the phase derivative is equal to zero. At these points $\phi \approx 0, \pi$. The dependencies of $\dot{\phi}_i$ on the initial phase $\phi_0$ and $\phi_i$ on time are shown in Fig. 6. One can see that, if the phases of dipoles are close to the point of repulsion, $\phi_r \approx 0$, then they diverge approaching $\phi_r$, whereas the phases that are close to the point of attraction, $\phi_a \approx \pi$, converge to $\phi_a$ with time.

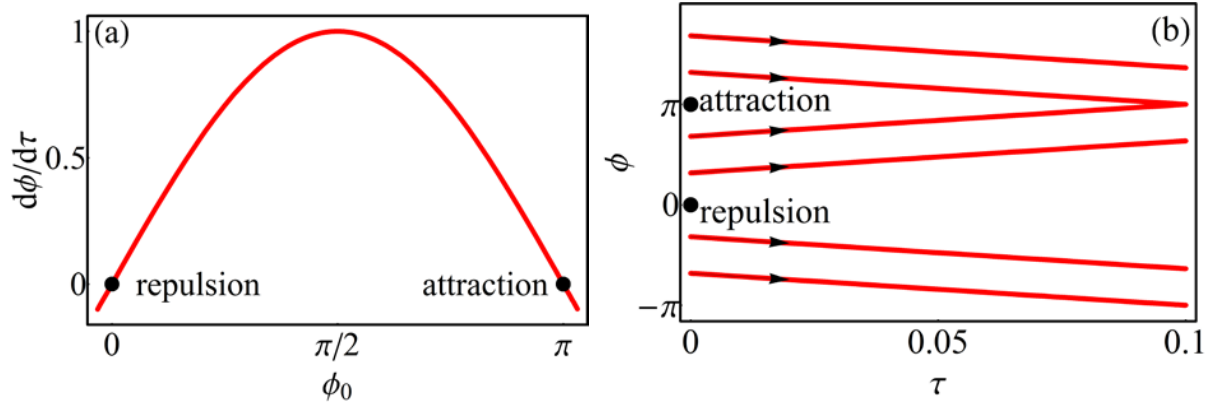

FIG. 6. (a) The dependence $\dot{\phi}_i$ on the initial phase $\phi_0$ and (b) the phase dynamics for small times given by the equation $\dot{\phi}_i = E_0 \sin \phi_i$.

This analysis is in a qualitative agreement with the computer simulation of the MF dynamics of phases $\phi_i$ shown in Fig. 7. There are two fixed points, an attraction point $\phi_a = 0.49\pi$ and repulsion point $\phi_r = -0.25\pi$. The phases of envelopes of some dipole oscillations draw close forming a "time speckle" at the attraction point $\phi_a \approx 0.49\pi$. At this moment, constructive interference in these envelopes occurs. Moreover, high-frequency oscillations interfere constructively as well because, as is noted above, all fast oscillations are in phase.

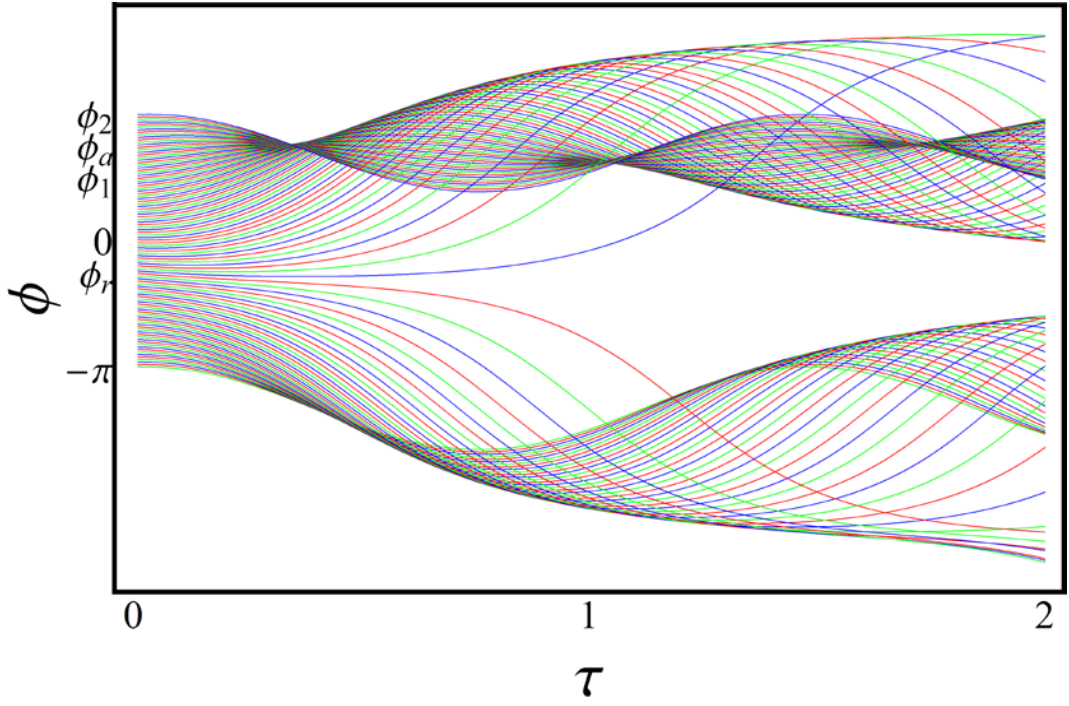

FIG. 7. The dynamics of phase oscillations of the envelops for non-interacting dipoles with $E \neq 0$. $\phi_a = 0.49\pi$, $\phi_r = -0.25\pi$, $\phi_1 = 0.49\pi$ and $\phi_2 = 0.97\pi$.

Comparing Figs. 5 and 7 we see that the phase condensation points and the SR peaks happen at the same times. This leads to the conclusion that SR pulses arise due to the constructive interference in the envelopes $c_i(\tau)$ of dipoles which initial phases belong to the interval $(\phi_1, \phi_2)$. We need to estimate the size of this interval and the lifetime of the SR pulse.

The lifetime of the phase fluctuation is of the order of the period of envelope oscillations, $\tau_{env}$, which is much greater than the main period of dipole oscillations. In Fig. 5, $\tau_{env} \sim 5$ while the period of dipole oscillations is $\tau_d \sim 1/(\omega \tau_N) \approx 10^{-3}$. During $\tau_{env}$, emitters have enough time to superradiate and to lose their energy. We emphasize that not all dipoles take part in the SR shown in Fig. 5. Obviously, the phenomenon is more pronounced if the number of dipoles with initial phases in the interval $(\phi_1, \phi_2)$ increases.

Let us estimate the size of the fluctuation forming a SR peak. During the time $\tau_0$, when this peak is formed, the phases of a maximum number of dipoles must converge. Since $A_i \sim 1$, then from Eq. (17) one can obtain an estimate for an optimal size of the initial phase fluctuation

$$\Delta \phi \sim E_0 \tau_0. \tag{18}$$

For a greater value of $\Delta\varphi$, phases do not have enough time to converge. For smaller $\Delta\varphi$, the number of dipoles taking part in the SR decreases. Fluctuations much smaller than the optimal one do not affect the system dynamics, because the energy that they radiate is small and the time of radiation is much longer than the lifetime of the SR fluctuation. The numerical calculations show that the intensity the radiation, $I = \left|\sum_i c_i\right|^2$, depends quadratically on the number of dipoles participating in the SR fluctuation (see Fig. 8).

Let us now estimate the delay time of the first SR peak. In both Dicke[1] and Vainshtein-Kleev[10] models, the dimensionless delay, $\tau_0$, is proportional to $\log N$. In the MF model described here, the delay time weakly depends on the number of dipoles forming the SR fluctuation [i.e., the number of equation in system (16)]. As a result, the delay time does not depend on the number of dipoles in the system. However, one needs to take into account that the MF in our model depends on the number of dipoles in the system. The numerical experiment shows (Fig. 8) that the delay time depends on the MF as $\tau_0 \sim (E_0)^{-\alpha}$, $\alpha \approx 0.5$. Thus, the greater the MF is, the faster emitters are phased. Assuming that $E_0 \sim (N)^{1/2}$ we obtain $\tau_0 \sim (N)^{-\alpha/2} \sim N^{-0.25}$. This dependence is close to logarithmic.

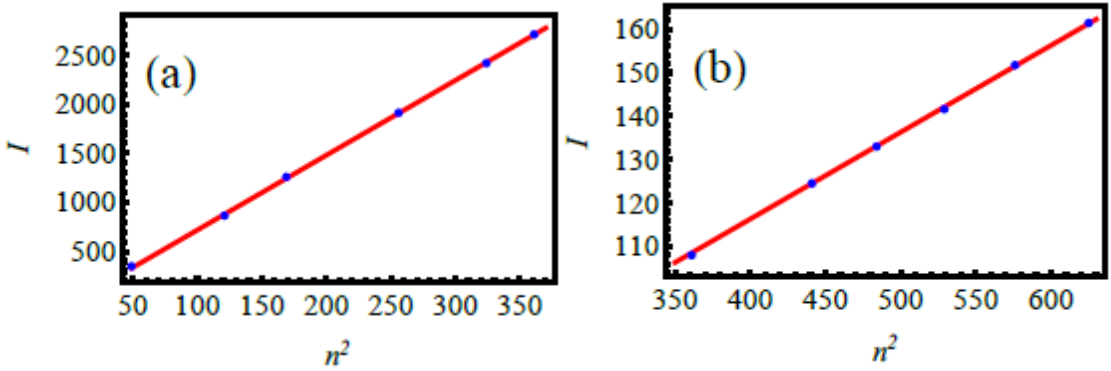

FIG. 8. The dependence of the oscillation intensity on the squared number of dipoles forming the SR fluctuation in (*a*) the MF model and (*b*) the exact model of Ref. 10.

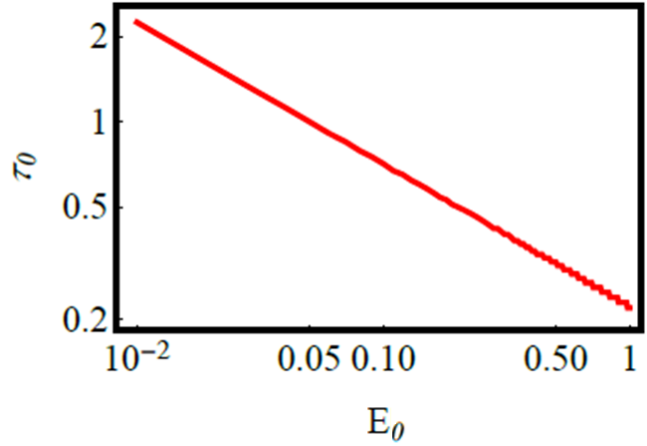

FIG. 9. The dependence of the delay time $\tau_0$ on the MF $E_0$.

In summary, the MF model shows that there is an "optimal" number of dipoles participating in forming a SR peak. Initially, the phases of these dipoles are spread in the interval defined by Eq. (18). As a result, the time delay of the SR peak does not depend on the total number of dipoles. In the exact model,[10] this dependence is weak (logarithmic) which may be related to the probability of forming the "optimal" fluctuation.

The main feature of the MF model is also observed in the exact model. As mentioned above, if the initial phase distribution is uniform, then the field of the radiative response is zero and there is no SR. If there is an initial fluctuation of phases in the system [the trajectories of dipoles forming the fluctuation are shown by blue lines in Fig. 10 (b)], a SR peak arises simultaneously with the constructive interference in envelops of the fluctuation (Fig. 10). Moreover, similarly to the exact model, in the MF model, the intensity of the SR peak depends not on the total number of dipoles in the system, but on the number of dipoles forming the fluctuation (see Fig. 8).

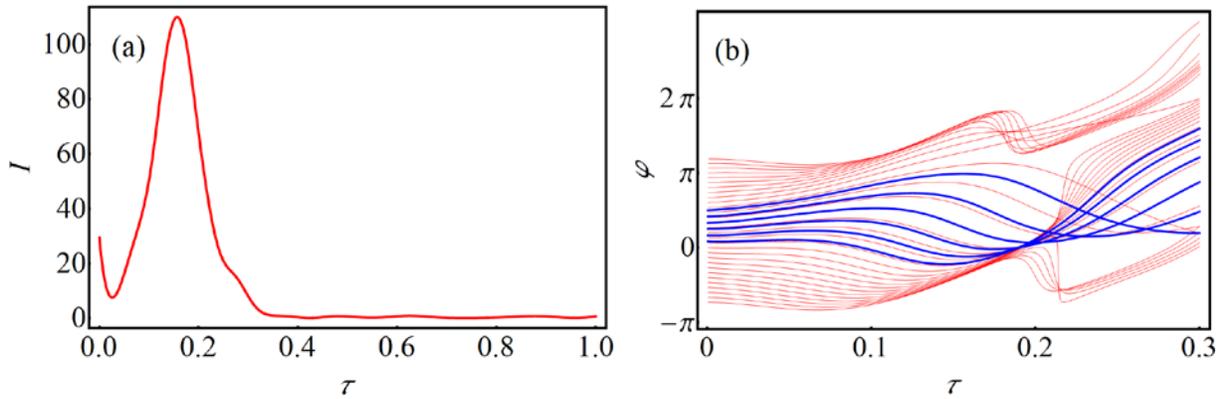

FIG. 10 (a) The dependence of the SR pulse intensity on the time and (b) the time dynamics of dipole phases with a non-zero initial fluctuation; blue lines denote trajectories of phases of dipoles forming the fluctuation.

## V. CONCLUSION

The suggested mechanism for the formation of a SR burst is based on the possibility of a fluctuation in the distribution of initial phases of dipoles. If phases of nonlinear dipoles are distributed uniformly in the interval $(-\pi,\pi)$, i.e. the number of dipoles $dN(\varphi)$ having the phase $\varphi$ in the interval $d\varphi$ is $Nd\varphi/2\pi$, then SR does not occur. If there is a finite interval of phases $(\varphi_1,\varphi_1)$ in which the number of dipoles $N_{\varphi_1\varphi_2}$ is greater than $N(\varphi_2-\varphi_1)/2\pi$, then over the time $t_0 \sim ln\,N_{\varphi_1\varphi_2}$, a SR peak with the intensity $\sim N_{\varphi_1\varphi_2}^2$ arises. In particular, this explains the SR peak in a system of dipoles with a random distribution of initial phases. Thus, the SR observed earlier in numerical experiments on the dynamics of a subwavelength array of classical nonlinear dipoles[10] is a purely classical phenomenon. It is not necessary to assume that emitters are identical. Note that for linear oscillators such a mechanism would not work because the dipole phase depends linearly on an external force. If a phase fluctuation arises, its lifetime would be of the same order as the period of fast oscillations of an emitter. This cannot result in noticeable radiation of the field.

Thus, the SR arises due to a low-frequency modulation of oscillations of a nonlinear dipole acted upon by a field of neighbors. The frequency of the modulation depends on both the initial phase of the oscillator and the total near-field of radiation. The SR peak is a result of the constructive interference in slow envelopes of the dipole oscillations. SR arises when dipole phases coincide. The duration of the SR burst is determined by the frequency of envelopes of the fast dipole oscillations.

## ACKNOWLEDGEMENTS

The work was supported by RFBR grants No. 13-02-92660, by Dynasty Foundation, by the NSF under Grant No. DMR-1312707, and by PSC-CUNY research award.